\documentclass[aps,pre,amsmath,amssymb,showpacs,showkeys,unsortedaddress]{revtex4}
\usepackage{graphicx}
\usepackage{dcolumn}
\usepackage{bm}

\newcommand{\la}{\left\langle}         
\newcommand{\ra}{\right\rangle}        
\newcommand{\mr}[1]{{\mathrm {#1}}}    
\newcommand{\kB}{k_\mr{B}}             
\newcommand{\Eq}[1]{Eq.~(\ref{#1})}    
\newcommand{\Eqs}[1]{Eqs.~(\ref{#1})}  
\newcommand{\Rf}[1]{Ref.~\cite{#1}}    
\newcommand{\Rfs}[1]{Refs.~\cite{#1}}  
\newcommand{\Fg}[1]{Fig.~\ref{#1}}     
\newcommand{\Sc}[1]{Sec.~\ref{#1}}     
\begin{document}

\title{ Influence of saving propensity on the power law tail of wealth distribution }

\author{Marco Patriarca}
  \email{marco.patriarca@mac.com}
  \homepage{http://staff.uni-marburg/~patriarc}
  \affiliation{Fachbereich Chemie, Philipps-Universit\"at Marburg,
  35032 Marburg, Germany}

\author{Anirban Chakraborti}
  \email{anirban@bnl.gov}
  \homepage{http://www.cmth.bnl.gov/~anirban/}
  \affiliation{Brookhaven National Laboratory,
  Department of Physics, Upton, New York 11973, USA}

\author{ Guido Germano }
  \email{germano@staff.uni-marburg.de}
  \homepage{http://staff.uni-marburg/~germano}
  \affiliation{Fachbereich Chemie, Philipps-Universit\"at Marburg,
  35032 Marburg, Germany}

\date{\today}

\begin{abstract}

Some general features of kinetic multi-agent models are reviewed, with particular attention to the relation between the agent saving propensities and the form of the equilibrium wealth distribution. The effect of a finite cutoff of the saving propensity distribution on the corresponding wealth distribution is studied. Various results about kinetic multi-agent models are collected and used to construct a realistic wealth distribution with zero limit for small values of wealth, an exponential form at intermediate and a power law tail at larger values of wealth.

\end{abstract}

\maketitle

\graphicspath{{./fig/}}  


\section{Introduction}

Multi-agent models of closed economy systems have received considerable attention in recent years  due to the fact that they seem to predict realistic shapes of wealth distribution from very simple underlying dynamics, basically equivalent to kinetic theory of ideal gases in classical statistical mechanics \cite{Bennati1988a,Bennati1988b,Bennati1993a,Ispolatov1998a,Dragulescu2000a,Chakraborti2000a,Chakraborti2002a,Chatterjee2003a,Iglesias2004a,Chatterjee2004a,Repetowicz2004a,Patriarca2004b,Patriarca2004c,Chatterjee2005a,Patriarca2005d}.
A notable plus-point of these simple models is represented by the ability to reproduce the main features of the empirical wealth distributions: a Boltzmann distribution at intermediate values of wealth, and a power law at the highest values (see e.g. \cite{Dragulescu2001a,Dragulescu2001b,Fujiwara2003a,Levy1997a,Sinha2005a}). The power law form in the tail of the distribution was observed more than a century ago by the economist Vilfredo Pareto \cite{Pareto1897a}, who found that the wealth of individuals in a stable economy has a cumulative distribution $F(x)\propto x^{-\alpha }$, where $\alpha$, the Pareto exponent, has usually a value between $1$ and $2$.

In these models $N$ agents interact exchanging a quantity $x$, which can be interpreted as representing any economic entity contributing to the agent wealth, expressed in the same unit of measure, e.g. in monetary units.
Depending on the parameters of the kinetic model, in particular on the values of the saving propensities $\{\lambda_i\}$ ($i=1,\dots,N$) of the $N$ agents, the equilibrium wealth distribution can be a simple Boltzmann distribution for $\lambda_i \equiv 0$ \cite{Bennati1988a,Bennati1988b,Dragulescu2000a}, a Gamma distribution with a similar exponential tail but a well defined mode $x_\mr{m}>0$ for $\lambda_i \equiv \lambda_0 > 0$ \cite{Chakraborti2000a,Chakraborti2002a}, or a distribution with a power law tail for randomly distributed $\lambda_i$ \cite{Chatterjee2003a,Chatterjee2004a}.
It has been recently recognized \cite{Chatterjee2004a,Patriarca2005d} that the observed power law arises from the overlap of Gamma distributions, resulting from (subsets of) agents with similar values of $\lambda$.
That is, in systems where saving propensity is distributed according to an arbitrary distribution function $f(\lambda$), agents relax individually toward Maxwell-Boltzmann distributions, similarly to systems with a global saving propensity $\lambda_0$, but with the important difference that in this case the various Gamma distributions with different $\lambda_0$ parameters will overlap and provide the final (power law) equilibrium distribution.

The aim of the present paper is to further investigate the relation between the saving propensity distribution and the shape of the final equilibrium wealth distribution, with particular attention to reproduce a realistic distribution.
In \Sc{ model } we recall the main features of kinetic multi-agent models, while in \Sc{ real } we consider how the equilibrium distribution is affected by a particular choice of the parameters of the saving propensity distribution and provide some examples. Results are summarized in \Sc{ conclusions }.


\section{ Kinetic multi-agent models }
\label{ model }

In kinetic multi-agent models $N$ agents interact with each other through a pair interaction -- but this is only one of the many possibilities -- exchanging a quantity $x$, generally referred to as ``wealth'' in the following.
Agents are characterized by their current wealths $\{x_i\},~i=1,2,\dots,N$ and, possibly, by some parameters, such as the saving propensity $\lambda_i$.
The evolution of the system is then carried out in the following way.
At every time step two agents $i$ and $j$ are extracted randomly and an amount of wealth $\Delta x$ is exchanged between them, 
\begin{eqnarray}
  x_i' &=& x_i - \Delta x \, ,
  \nonumber \\
  x_j' &=& x_j + \Delta x \, .
  \label{basic0}
\end{eqnarray}
It can be noticed that in this way the quantity $x$ is conserved during the single transactions, $x_i' + x_j' = x_i + x_j$, where $x_i'$ and $x_j'$ are the agent wealths after the transaction has taken place.

\subsubsection{ The basic model }
\label{ basic }

In the basic versions of the model the quantity $x$ represents money and $\Delta x$ the money exchanged, assumed to have a constant value 
\cite{Bennati1988a,Bennati1988b,Bennati1993a},
\begin{equation}
  \Delta x = \Delta x_0 \ ,
\end{equation}
or to be proportional to the initial values \cite{Dragulescu2000a},
\begin{equation} \label{Dx2}
  \Delta x =  \bar{\epsilon} x_i - \epsilon x_j \, ,
\end{equation}
where $\epsilon$ is a random number uniformly distributed between 0 and 1 and $\bar{\epsilon}  = 1 - \epsilon$.
The form of $\Delta x$ in \Eq{Dx2} represents a random reshuffling of the wealths of the two agents \cite{Dragulescu2000a}, 
since \Eq{basic0} can in this case be rewritten as
\begin{eqnarray}
  x_i' &=& \epsilon (x_i + x_j) \, ,
  \nonumber \\
  x_j' &=& \bar{\epsilon} (x_i + x_j) \, .
  \label{basic1}
\end{eqnarray}
These dynamics rules, together with the constraint that transactions can take place only if $x_i'>0$ and $x_j'>0$,
lead to an equilibrium state characterized by an exponential Boltzmann distribution,
\begin{equation}\label{f-basic}
  f(x) = \la x \ra^{-1} \exp(-x/\la x \ra) \, ,
\end{equation}
where the effective temperature $T_\lambda$ of the system is just the average wealth (see curve $\lambda=0$ in \Fg{fig:gamma}).
\begin{figure}[tb] 
  \begin{center}
    \includegraphics[angle=0,width=.45\textwidth]{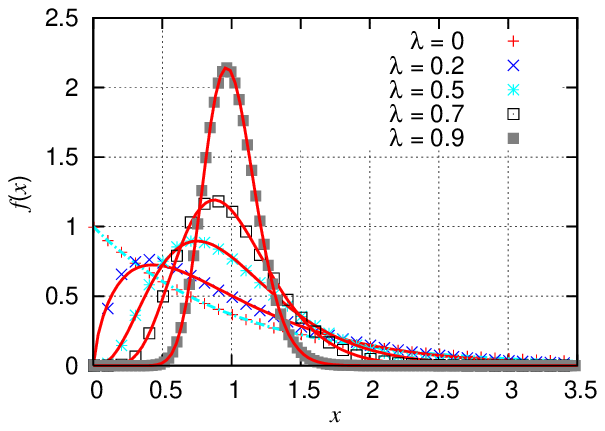}
    \includegraphics[angle=0,width=.45\textwidth]{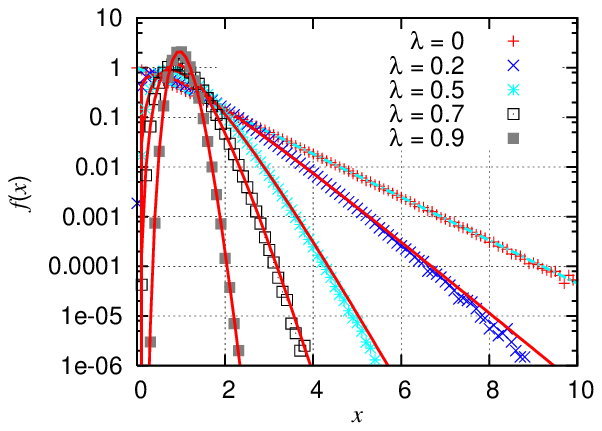}
    \caption{ Linear (left) and semi-log (right) plots of the probability density of wealth $x$ from numerical simulations (dots) for various values of the global saving propensity $\lambda$ in the interval (0,1), compared with the theoretical curves (continuous lines) defined by \Eqs{f-global}, (\ref{xi}), and (\ref{DT}). The curve for $\lambda=0$ is the Boltzmann distribution.} 
    \label{fig:gamma} 
  \end{center}
\end{figure} 
Despite its intrinsic simplicity, the basic model has the merit of having shown that economic interactions can be modeled in terms of simple statistical mechanisms leading to corresponding universal statistical laws. 

\subsubsection{ Models with a global saving propensity }
\label{ global }

A first generalization toward a more realistic model is based on the introduction of a saving criterion.
Agents save a fraction $\lambda$ (the saving propensity, with $0 < \lambda < 1$) before entering a trade and only exchange the remaining fraction $(1-\lambda)$ of their wealth, \cite{Chakraborti2000a,Chakraborti2002a}:
\begin{eqnarray}
  x_i' &=& \lambda x_i + \epsilon (1-\lambda) (x_i + x_j) \, ,
  \nonumber \\
  x_j' &=& \lambda x_j + \bar{\epsilon} (1-\lambda) (x_i + x_j) \, ,
  \label{sp1}
\end{eqnarray}
corresponding to a $\Delta x$ in \Eq{basic0} given by
\begin{equation}
  \Delta x  
  =     (1 - \lambda) [ \bar{\epsilon} x_i - \epsilon x_j ]  \, .
\end{equation}
The corresponding equilibrium distribution is well fitted by the gamma distribution \cite{Patriarca2004b,Patriarca2004c}
\begin{equation}\label{f-global}
  f(\xi) 
  = \frac{1}{\Gamma(D_\lambda/2)} \, \xi^{D_\lambda/2-1} \exp( - \xi )
  \equiv \gamma_{D_\lambda/2}(\xi) \, ,
\end{equation}
as shown in \Fg{fig:gamma}. Here the dimensionless variable
\begin{equation}
  \xi = {x \over T_\lambda} \, ,
  \label{xi}
\end{equation}
is just the variable $x$ rescaled with respect to the effective temperature $T_\lambda$ and
\begin{eqnarray}
  \frac{D_\lambda}{2} 
  &=& 1 + \frac{3 \lambda}{1 - \lambda} = \frac{1 + 2\lambda}{1 - \lambda} \, ,
  \nonumber \\
  T_\lambda 
  &=& \frac{1 - \lambda}{1 + 2\lambda}\la x \ra \, .
  \label{DT}
\end{eqnarray}
The parameter $D_\lambda$ plays the role of an effective dimension, since the Gamma distribution $\gamma_{n}(\xi)$ given by \Eq{f-global} is identical to the Maxwell-Boltzmann distribution of kinetic energy for a system of molecules at temperature $T_\lambda$ in $D_\lambda$ dimensions (of course only for integer or half-integer values of $n=D_\lambda/2$) \cite{Patriarca2004c,Patriarca2005d}.
In further support of this analogy, it is worth noting that $T_\lambda$ and $D_\lambda$ are related to each other through an ``equipartition theorem'',
\begin{equation}
  \la x \ra = { {{D_\lambda} T_\lambda} \over 2} \, .
  \label{equipartition}
\end{equation}
The equivalence between kinetic theory and closed economy models, suggested by the basic version of the kinetic multi-agent models \cite{Bennati1988a,Bennati1988b,Bennati1993a,Dragulescu2000a}, can thus be extended to values $\lambda \ge 0$ \cite{Patriarca2005d}, as summarized in Table \ref{tab:analogy}.
\begin{table}
\centering
\caption{Analogy between kinetic and multi-agent model}
\label{tab:analogy}
\begin{tabular}{lll}
\hline\noalign{\smallskip}
                & Kinetic model       & Economy model  \\
\noalign{\smallskip}\hline\noalign{\smallskip}
variable        & $K$=kinetic energy  & $x$=wealth \\
units           &  $N$ particles      & $N$ agents\\
interaction     & collisions          & trades\\
dimension       & integer $D$         & real number $D_\lambda$\\
equipartition theorem
                & $\kB T=2\la K \ra/D$& $T_\lambda=2\la x \ra/D_\lambda$\\
reduced variable& $\xi = K/\kB T$     & $\xi = x / T_\lambda$\\
distribution    & $f(\xi)=\gamma_{D/2}(\xi)$ & $f(\xi)=\gamma_{D_\lambda/2}(\xi)$\\
\noalign{\smallskip}
\hline
\end{tabular}
\end{table}

While $\lambda$ varies between 0 and 1, the effective dimension $D_\lambda$ increases monotonically between 2 and $\infty$.
In fact in a higher number of dimensions the fraction of kinetic energy exchanged between particles during a collision is smaller.
At the same time, the market temperature $T_\lambda$ decreases with increasing $\lambda$, signaling smaller fluctuations of $x$ during trades, consistently with the presence of a saving criterion, i.e. $\lambda>0$.
One can notice that $T_\lambda = (1-\lambda)\la x \ra/(1+2\lambda) \approx (1-\lambda)\la x \ra$ is on average the amount of wealth exchanged during an interaction between agents, see \Eqs{sp1}.

\subsubsection{ Models with a continuous distributions of saving propensity }
\label {continuous}

\begin{figure}[tb]
  \begin{center}
    \includegraphics[angle=0,width=0.32\textwidth]{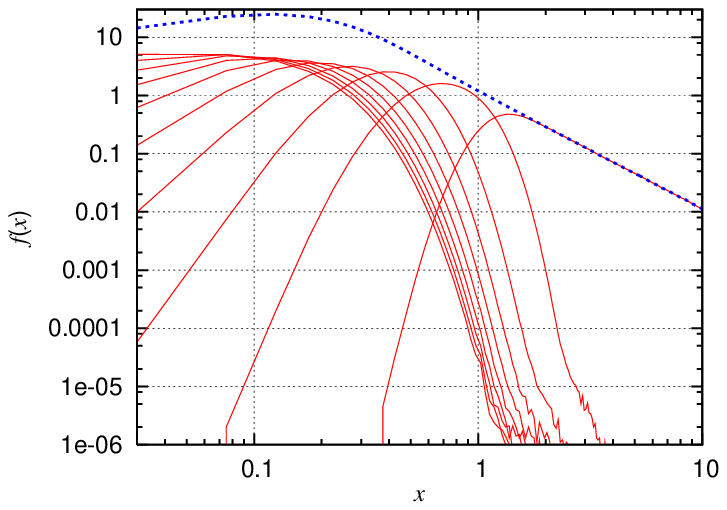}
    \includegraphics[angle=0,width=0.32\textwidth]{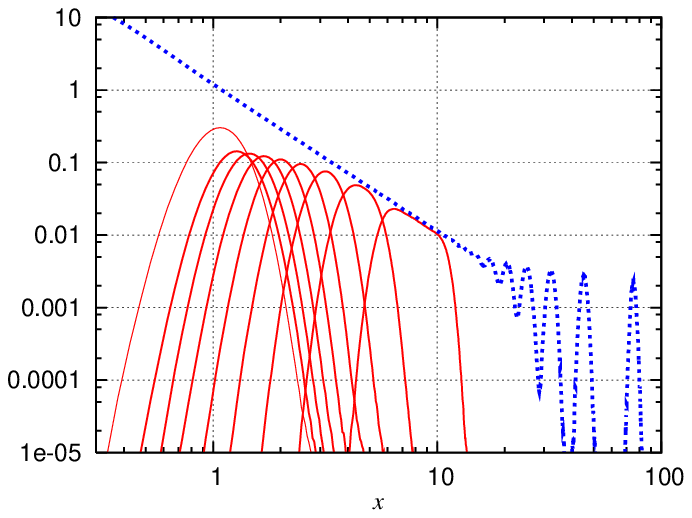}
    \includegraphics[angle=0,width=0.32\textwidth]{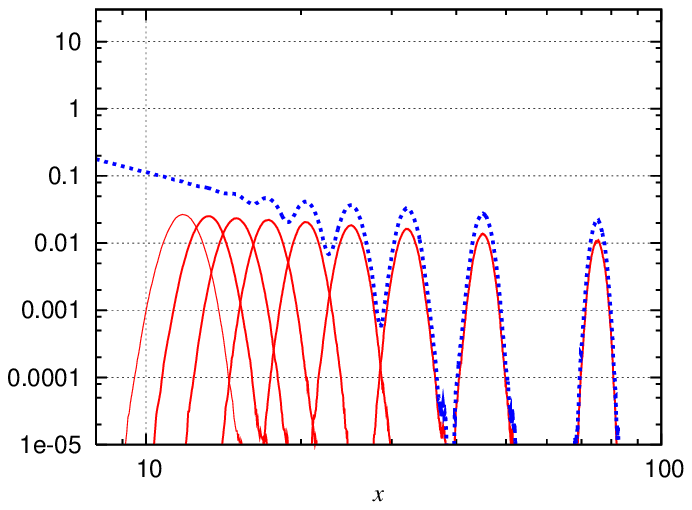}
    \caption{
    Log-log plot of wealth distribution (dotted line) resolved into partial distributions (continuous curves) corresponding to agents with a saving propensity belonging to a sub-interval of the $\lambda$ range (0,1). Left: Partial distributions from the 10 intervals of width $\Delta\lambda=0.1$ of the $\lambda$ range (0,1). Center: Partial distributions from the interval $\lambda=(0.9,1.0)$ further resolved into partial distributions from sub-intervals of width $\Delta\lambda=0.01$. Right: Partial distribution from the interval $\lambda=(0.99,1.00)$ -- not shown in the central figure -- resolved into partial distributions from intervals of width $\Delta\lambda=0.001$. Peaks are due to the power law breaking down as the distance between consecutive partial distributions becomes comparable with their width.
    }
    \label{partials_a}
  \end{center}
\end{figure}
As a further generalization, various investigations concerned models in which agents have realistically been diversified from each other by assigning them different saving propensities $\lambda_i$ \cite{Chatterjee2003a,Das2003a,Chatterjee2004a,Repetowicz2004a,Chatterjee2005a,Patriarca2005d}.
In particular, uniformly distributed $\lambda_i$ in the interval (0,1) have been studied numerically in \Rfs{Chatterjee2003a,Chatterjee2004a}.
This model is described by the trading rule 
\begin{eqnarray}
  x_i' &=& 
  \lambda_i x_i + \epsilon [ (1-\lambda_i) x_i + (1-\lambda_j) x_j ] \, ,
  \nonumber \\
  x_j' &=& 
  \lambda x_j + \bar{\epsilon} [(1-\lambda_i) x_i + (1-\lambda_j) x_j ] \, ,
  \label{sp2}
\end{eqnarray}
or, equivalently, by a $\Delta x$ -- as defined in \Eq{basic0} -- given by
\begin{equation}
  \Delta x  
  =  \bar{\epsilon} (1-\lambda_i) x_i - \epsilon (1-\lambda_j) x_j  \, .
\end{equation}
One of the main features of this model, which is supported by theoretical considerations \cite{Das2003a,Repetowicz2004a,Chatterjee2005a}, is that the wealth distribution exhibits a robust power law at large values of $x$, 
\begin{equation}\label{f-power}
  f(x) = x^{-\alpha - 1} \, ,
\end{equation}
with a Pareto exponent $\alpha=1$ largely independent of the details of the $\lambda_i$-distribution.
As remarked in \Rf{Chatterjee2004a}, the wealth distribution of the single agents are not of a power law type but have a well defined mode and an exponential tail, similarly to the case a global saving propensity $\lambda_0$. The power law actually results from the overlap of these partial distributions corresponding to the various $\lambda$'s, which are Gamma distributions, whose average value is proportional to $1/(1-\lambda)$  and thus extend to very large values of $x$ \cite{Patriarca2005d}. These results are also in agreement with theoretical approaches to kinetic multi-agent models \cite{Repetowicz2004a,Das2003a,Chatterjee2005a}.
This phenomenon is illustrated in \Fg{partials_a}.

\begin{figure}[tb]
  \begin{center}
    \includegraphics[angle=0,width=0.45\textwidth]{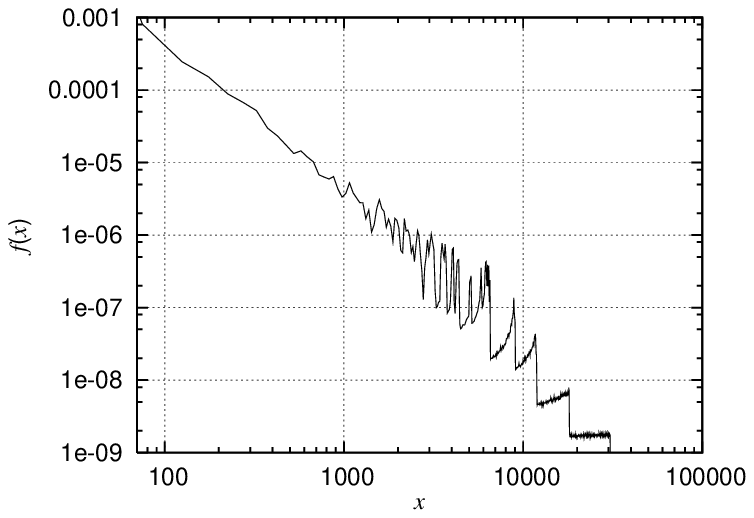}
    \includegraphics[angle=0,width=0.45\textwidth]{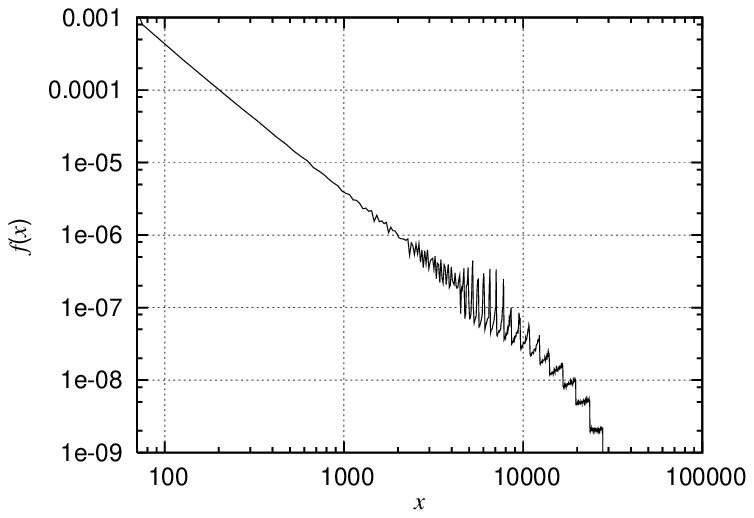}
    \caption{
    Reproduced from \Rf{Patriarca2005d}: Wealth distribution of a system of $N=10^6$ agents after $10^{12}$ trades for two uniform distributions of $\lambda$: A randomly generated $\lambda$-distribution (left) produces a wealth distribution which is more irregular than a deterministic uniform $\lambda$ distribution (right) in which $\lambda_i = i/N$.}
    \label{Ldistr}
  \end{center}
\end{figure}
Arbitrarily small (random) irregularities in the distance between two consecutive values of $\lambda$ close enough to 1, in a uniform distribution of saving propensity, are amplified in the wealth distribution as a consequence of the correlation between average wealth and saving propensity, resulting in isolated peaks in the wealth distribution \cite{Patriarca2005d}. This is shown by the simple example in \Fg{Ldistr}, where two distributions, in principle equivalent to each other since associated to a uniform $\lambda$-distribution in (0,1), look actually quite different: The first distribution (left) has been obtained by randomly extracting the values of $\lambda$ with a random number generator in the interval (0,1). It can be noticed that the corresponding equilibrium wealth distribution is more irregular than that obtained from the second distribution (right), in which the values $\lambda_i$ were set to be equidistant from each other in the interval (0,1) by defining them as $\lambda_i = i/N$.
The deterministic and random uniform distributions are equivalent to each other in principle but not in practice (within numerical simulations), where a finite number of agents is necessarily used. The reason is that the $\lambda$ values extracted randomly present fluctuations and therefore wider intervals between neighbor values which are amplified in the final wealth distribution. Since in these numerical simulations one tries to mimic continuous distributions by use of the smallest possible number of variables, it is convenient to avoid the irregular fluctuations present in randomly extracted sets of numbers and use a deterministically extracted sets $\{\lambda_i\}$ of saving propensities. This can be achieved easily with the method prescribed in Appendix A.


\section{ Construction of a realistic model }
\label { real }

Here we study how some aspects of the $\lambda$-distribution $f(\lambda)$ influence the equilibrium form of the wealth distribution $f(x)$, i.e. its shape at smaller and the tail at larger values of $x$, and in particular under which conditions an exponential and a power law can appear in different ranges of the same distribution. 

\subsection{ Wealth distribution at small and intermediate values of wealth }

The equilibrium distribution of the basic model is the simple exponential function in Eq. (\ref{f-basic}). Such a form of distribution decreases monotonously with $x$ and does not have rich agents nor a power law tail, a point dealt with in greater detail in the next section. 
In the small $x$ limit, the exponential distribution is $f(x \to 0) >0$  which implies that many agents have a wealth $x \approx 0$. In fact the mode of the distribution is $x_\mr{m} = 0$ and the fraction of agents outside a given interval $(0,x)$ -- which is just the upper cumulative distribution function -- has a pure exponential form, $F(x) = \exp(-x/\la x\ra)$. Real data about wealth and income distributions, on the other hand, show that wealth distribution functions have a mode $x_\mr{m} > 0$ \cite{Dragulescu2001b,SalaiMartin2002a,SalaiMartin2002b,Aoyama2003a,Ferrero2004a,Silva2005a}. 
The introduction of a (global) saving propensity $\lambda>0$ solves this problem \cite{Chakraborti2000a} since it leads to an equilibrium Gamma distribution \cite{Patriarca2004b,Patriarca2004c}, which has a mode $x_m>0$ and a zero limit for $x \to 0$, see \Fg{fig:gamma}. 
This functional form has been shown to interpolate well real data about income distributions at small and intermediate values \cite{Dragulescu2001a,Dragulescu2001b,Ferrero2004a,Silva2005a}.

\subsection{ The tail of the wealth distribution }

The tail of wealth distributions is known to follow a power law with Pareto exponent between 1 and 2, depending on the sample analyzed.
The model under consideration, when saving propensities are continuously distributed, predicts a power law tail in $f(x)$, despite with a lower Pareto exponent $\alpha=1$, a feature which has been shown to be very robust and independent of the details of $f(\lambda)$. 

Both numerical and theoretical analyzes of kinetic multi-agent models show that agents with large values of $\lambda$'s (i.e. $\lambda$ close to 1) give a major contribution to the power law tail.
This is illustrated e.g. by the fact that when the power law is decomposed into partial distributions of agents within a given interval of the saving propensity, the partial distribution corresponding to the interval with the highest $\lambda$ is in turn a power law, while the distributions corresponding to lower values of $\lambda$ are localized and have an exponential tail, as shown in \Fg{partials_a}. However, they sum up to give a power law at lower values of $x$. All this suggests that the crucial factor for having a power law extending beyond a certain value $x$ is the highest $\lambda$ present in the sample, that is the cutoff of the $\lambda$-distribution.

Thus, rather than varying the functional form of $f(\lambda)$, the influence of the cut-off  $\lambda_\mr{M}$ of the $\lambda$-distribution -- which is a parameter characterizing numerical simulations as well as real systems -- has been analyzed.
A uniform deterministic distribution of saving propensity for the $N$ agents in the interval ($0,\lambda_\mr{M}$), has been generated through the formula
\begin{equation}\label{lambdamax}
  \lambda_i 
  = \left(\frac{i}{N}\right) \lambda_\mr{M} \, ,~~~~i = 1,\dots,N,~~~~\lambda_\mr{M}<1 \, ,
\end{equation}
as described in greater detail in Appendix A.

In fact we found that varying the cutoff $\lambda_\mr{M}$ influences in turn the cut-off of the wealth distribution $f(x)$ and the shape of the distribution at small $x$ -- but not the shape of the tail which remains a power law with exponent $\alpha=1$.
Decreasing $\lambda_\mr{M}$ has the effect to decrease the interval of wealth $x$ in which the power law appears, until it eventually disappears for $\lambda_\mr{M} \approx 0.92$.
Results are shown in \Fg{cutoff}, where the various curves represent the distribution functions obtained for some values of the cutoff $\lambda_\mr{M}$ chosen in the interval $\lambda_\mr{M}=(0.9, 0.9999)$ in a system of $10^5$ agents. Curves from left to right correspond to increasing values of cutoff. 
\begin{figure}[tb]
  \begin{center}
    \includegraphics[angle=0,width=0.6\textwidth]{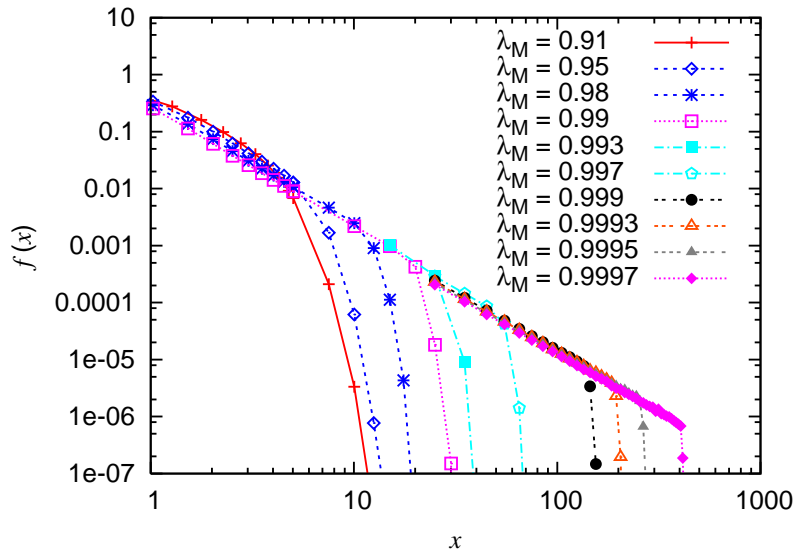}
    \caption{
    Wealth distribution obtained from a uniform saving propensities distributions $f(\lambda)$ defined by \Eq{lambdamax} in a system of $10^5$ agents for various values of the cutoff $\lambda_\mr{M}$. Curves from left to right correspond to increasing $\lambda_\mr{M}$.
    }
    \label{cutoff}
  \end{center}
\end{figure}
The transition from an exponential to a power law form of the wealth distribution, as the cut-off $\lambda_\mr{M}$ decreases, takes place by a shrinking of the power law interval, rather than as a change of the functional form of the tail.

As a final remark it is to be noted that the cutoff of the $\lambda$-distribution is naturally linked to that of the $x$-distribution, as a consequence of the correlation existing between average wealth $\la x \ra$ and saving propensity \cite{Patriarca2005d} in this model,
\begin{equation}\label{xL}
  \la x_i \ra (1-\lambda_i) = \mr{const} \, .
\end{equation}
Here the constant on the right hand side of the equation is the same number for all the agents in the system. This relation clearly shows that the highest average wealth is determined in turn by the highest $\lambda_i$.


\subsection{ Superposing an exponential form at intermediate values and a power law tail }

In real wealth distributions, an exponential form at intermediate values of wealth is known to coexist with a power law tail at larger values \cite{Silva2005a}.
The power law is mainly due to a small percentage of population, of the order of a few per cent, while the majority of the population with smaller average wealth give rise to the exponential part of the distribution.
In this section we try to construct a realistic example of such a type of wealth distribution by collecting some of the results obtained so far:
\begin{itemize}
\item
A global saving propensity $\lambda_0$ is associated to an equilibrium Gamma distribution, which always has an exponential tail.
\item
A set of agents with a continuous $\lambda$-distribution produces a power law in the equilibrium wealth distribution.
\item
The average wealth $\la x_i \ra$ of an agent and the corresponding saving propensity $\lambda_i$ are linked to each other through \Eq{xL}, which implies that agents with high $\lambda \approx 1$ contribute to the large-$x$ part of the distribution.
\end{itemize}
It is then natural to ask if a suitable $\lambda$-distribution may lead to the desired equilibrium wealth distribution. To answer this question we have constructed a hybrid $\lambda$-distribution -- on the base of the results listed above and a very similar prescription mentioned in \Rf{Chatterjee2003a}-- in the following way:
\begin{itemize}
\item
A small fraction of agents $p_0$ with saving propensities $\lambda_i$ uniformly distributed in the interval (0,1) according to \Eq{linear}.
\item 
The remaining fraction $1-p_0$ with a constant value of the saving propensity $\lambda_0$.
\end{itemize}

The corresponding distribution for $p_0=0.01$ (1 per cent) and $\lambda_0=0$ is shown in \Fg{mixed_L0}, both in the small $x$-scale, where the distribution has an exponential shape, and in the long $x$-range, where the power law with exponent $-2$, which characterizes this type of multi-agent model, is observed.
\begin{figure}[tb]
  \begin{center}
    \includegraphics[angle=0,width=0.48\textwidth]{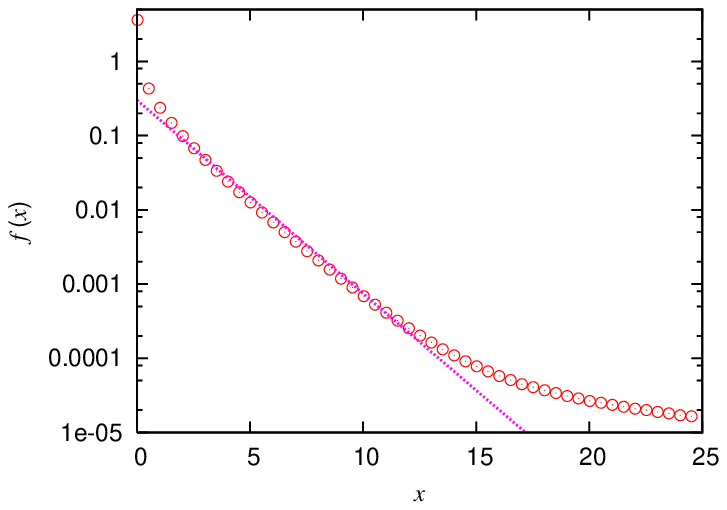}
    \includegraphics[angle=0,width=0.48\textwidth]{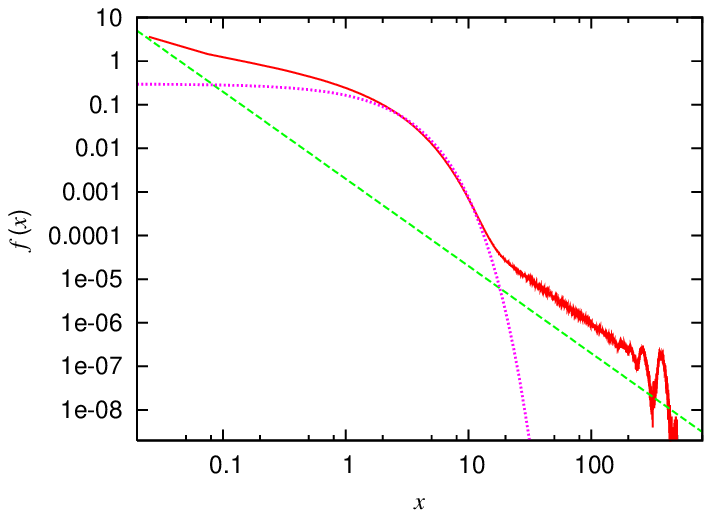}
    \caption{
    Numerical equilibrium wealth distribution of a population of $10^5$ agents. One per cent (1000) of the agents have uniformly distributed saving propensities in the interval (0,1), while the rest have $\lambda=0$. 
    Left (semi-log scale): Small $x$ part of the numerical distribution (dots) compared with an exponential function $\propto\exp(-3x/5) $(dotted line). 
    Right (log scale): Numerical distribution (continuous line) compared with a power law $\propto x^{-2}$ (dashed line) and the same exponential function of the left figure (dotted line). Peaks at high $x$ are due to the finite resolution in $\lambda$, see \Fg{partials_a} for an explanation of their origin.
    }
    \label{mixed_L0}
  \end{center}
\end{figure}

It is noteworthy that the coexistence of an exponential and a power law tail is possible only for small values of $p_0$, in agreement with the fact that it is a small percentage of the population in real systems that is responsible of the power law form of wealth distribution at large values of wealths. For larger values of $p_0$ the exponential part shrinks and the power law dominates.
This effect is in a sense contrary to that considered in \Sc{continuous}, where decreasing the cutoff of the $\lambda$-distribution induced a shrinking of the power law range.

\begin{figure}[tb]
  \begin{center}
    \includegraphics[angle=0,width=0.48\textwidth]{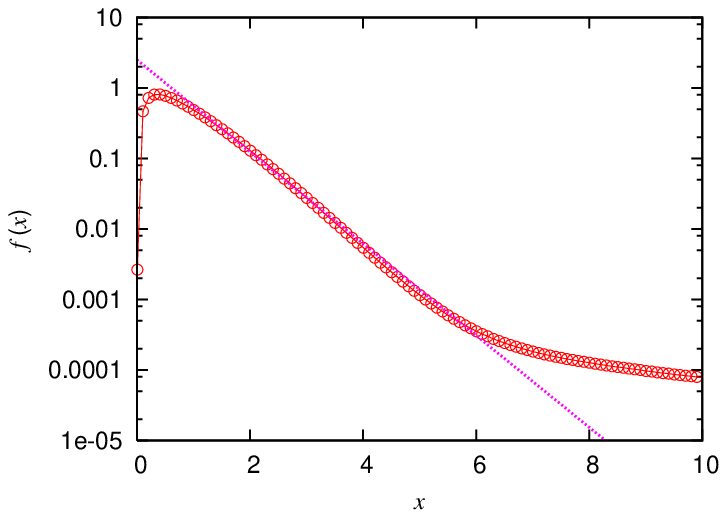}
    \includegraphics[angle=0,width=0.48\textwidth]{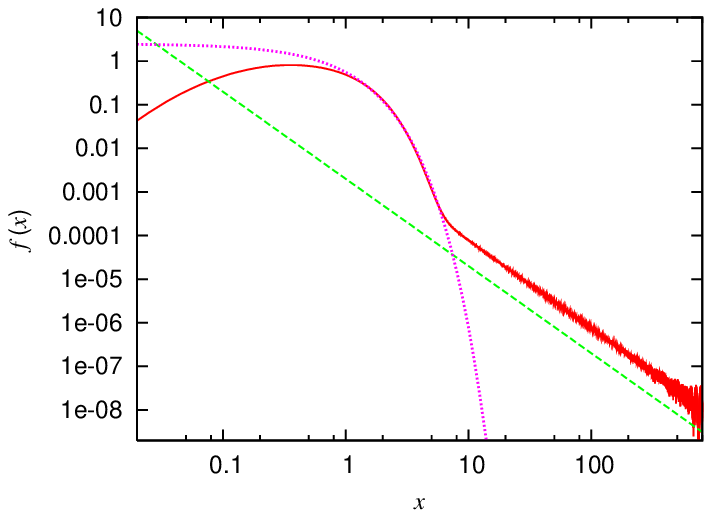}
    \caption{
    Numerical equilibrium wealth distribution of a population of $10^6$ agents, in which $10^4$ agents ($1\%$) have uniformly distributed saving propensities in the interval (0,1), while the remaining $9.9\times 10^5$ agents ($99\%$) have $\lambda=0.2$. Respect to \Fg{mixed_L0}, the distribution has a mode $x_\mr{m}>0$.
    Left (semi-log scale): Small scale part of the numerical distribution (dots-continuous line) compared with a function $\propto\exp(-3x/2)$ (dotted line). 
    Right (log scale): Numerical distribution (continuous line) compared with a power law $\propto x^{-2}$ (dashed line) and with the same exponential function of the left figure (dotted line).
    }
    \label{mixed_L02}
  \end{center}
\end{figure}
It may be noticed that, due to the choice $\lambda_0 = 0$ for that part of agents with a constant saving propensity, the distribution in \Fg{mixed_L0} still has a mode $x_\mr{m}=0$. However, one recovers a distribution with a well defined mode $x_\mr{m}>0$ as soon as one chooses a $\lambda_0 \ne 0$.
The distribution in \Fg{mixed_L02} corresponds to a $\lambda_0 = 0.2$ for 99\% of the agents and a uniform $\lambda$-distribution for the remaining agents.


\subsection{ Meaning of the saving propensity }

The central role of the saving propensity $\lambda$ -- or risk aversion as referred to in \Rf{Iglesias2004a} -- for the considerations made above is evident.
However, it is to be remarked that the relation between saving propensity $\lambda_i$ of an agent and the corresponding average wealth $\la x_i \ra$ should not considered to be of a cause-effect type.
It is true that in the present model the $\lambda_i$'s are fixed parameters, so that the natural dynamical interpretation is that the saving propensity $\lambda$ determines the final average wealth. However, in a real situation the value of $\lambda$ itself may vary according to various factors, e.g. the wealth itself: a high average wealth probably puts the agent in a situation which allows to carry on trades investing the same amount of wealth while saving more respect to agents with smaller wealths.
Therefore, the model contains in its very dynamics a positive correlation between $\lambda$ and $\la x \ra$ supported by real data \cite{Dynan2004a} but leaves the question of the actual dynamical relation between them to a more detailed microscopic analysis.
Multi-agents models like that considered here describe flux of wealth on a mesoscopic level, i.e. on a coarse grained scale in time or wealth, rather than reflecting the single agent strategy to save or reduce risks.


\section{ Conclusions }
\label{ conclusions }

We have shown that within the framework of kinetic multi-agent models it is possible to obtain realistic wealth distributions $f(x)$ characterized by a zero limit for small $x$, and the coexistence of an exponential form at intermediate and power law tail at larger values of $x$. In agreement with observations on real systems, this is possible only if the percentage of rich agents does not exceed a critical threshold of the order of 1 per cent. Also, the model naturally produces a positive correlation between average wealth $\la x \ra$ and saving propensity $\lambda$ exhibited in real data samples.


\begin{acknowledgments}
  Numerical computations were partially carried out on the facilities of the Laboratory of Computational Engineering, Helsinki University of Technology, under support by the Academy of Finland, Research Centre for Computational Science and Engineering, project no.~44897 (Finnish Centre for Excellence Program 2000-2005).
  The work at Brookhaven National Laboratory was carried out under Contract No. DE-AC02-98CH10886, Division
of Material Science, U.S. Department of Energy.
\end{acknowledgments}


\appendix

\section{ Extraction of a variable $\lambda$ with cumulative distribution $F(\lambda)$ }

It is possible to define a sequence of $N$ numbers $\lambda_i$, $i=1,\dots,N$, which becomes distributed according to an arbitrary distribution function $f(\lambda)=dF(\lambda)/d\lambda$ in the continuous limit ($N \to \infty$), in at least two ways, randomly or deterministically. The two methods are equivalent to each other only in the continuous limit, while in numerical simulations a finite $N$ is necessarily employed and they may provide different results. As discussed in \Sc{continuous}, in some cases it may be preferable to have a regular, rather than a randomly extracted sequence. 
\begin{itemize}
\item
{\em Random extraction}. A generator of random numbers $\phi$, $0<\phi<1$, uniformly distributed between 0 and 1, can be employed to extract a set of numbers $\lambda_i$ distributed in the continuous limit according to an arbitrary cumulative distribution function $F(\lambda)$, with $F(0)=0$ and $F(1)=1$. The cumulative distribution function for the random variable $\phi$ is simply $F(\phi) = \phi$ and $dF \equiv d\phi$ is the (constant) probability to extract the next random number between $\phi$ and $\phi+d\phi$. The algorithm is based on the identity $d\phi=dF=f(\lambda)d\lambda$, which shows that if values $F_i$ are extracted randomly and uniformly in the interval (0,1), then the corresponding values $\lambda_i$ obtained by inverting $F=F(\lambda_i)$ will be distributed with probability density $f(\lambda)$.
\item
{\em Deterministic extraction}
The same result can be obtained by a deterministic assignment of the values $\lambda_i$ which does not make use of random number generators.
If the sequence $\{\lambda_i\}$ is assumed to be labeled in increasing order, i.e. $0 \le \lambda_1 < \lambda_2 < \dots < \lambda_N \le 1$, then the function of $i$
\begin{equation}
  \lambda(i) = \lambda_i \, ,
\end{equation}
increases monotonously with $i$ and it is possible to invert it to express $i$ as a function of $\lambda_i$ to define the function 
\begin{equation}\label{F}
  F(\lambda_i) = \frac{i}{N} \, ,
\end{equation}
which represents the fraction of agents with saving propensity less or equal to $\lambda_i$: $F(\lambda)$ is just the (lower) cumulative distribution function and as such $0<F(\lambda)<1$ for every $\lambda$, $F(\lambda \to 0) \to 0$, and $F(\lambda \to 1) \to 1$. 
For instance the cumulative distribution function of a uniformly distributed variable $\lambda$ in the interval $\lambda \in (0,1)$ is just the linear function $F(\lambda) = \lambda$, with $0<\lambda<1$. Then \Eq{F} provides the corresponding deterministic sequence as
\begin{equation}\label{linear}
  \lambda_i = \frac{i}{N} \, ,~~~~~~~i=1,\dots,N \, . 
\end{equation}
If there is an upper cutoff $\lambda_\mr{M}$ in the distribution, the equation is modified as in (\ref{lambdamax}).
In the general case of a given cumulative function $F(\lambda)$, it is sufficient to invert \Eq{F} to obtain the sequence in the form $\lambda_i = \lambda(i/N)$, $i=1,\dots,N$, where $\lambda(\dots)$ is the inverse function of $F(\dots)$. The values $\lambda_i$ thus obtained will be distributed in the continuous limit with a probability distribution $f(\lambda)=dF(\lambda)/d\lambda$.
\end{itemize}

\bibliography{wehia}

\end{document}